\documentclass[aps,prl,twocolumn,groupedaddress,amsmath,superscriptaddress]{revtex4}

\usepackage{amsfonts}
\usepackage{amsmath}
\usepackage[dvips]{graphicx}
\usepackage{color}
\usepackage{amsmath}
\usepackage{graphicx}
\usepackage{amssymb}
\usepackage{hyperref}
\usepackage{color}
\usepackage{mathrsfs}
\usepackage{isomath}
\usepackage{amsmath}
\usepackage{amsthm}
\usepackage{epstopdf}
\usepackage{txfonts}
\usepackage{dsfont}

\makeatletter

\definecolor{nblue}{rgb}{0.3,0.3,1.0}
\definecolor{ngreen}{rgb}{0.2,0.7,0.2}
\definecolor{nred}{rgb}{0.9,0.1,0}
\definecolor{nblack}{rgb}{0,0,0}

\newcommand{\beq}{\begin{equation}}
	\newcommand{\eeq}{\end{equation}}
\newcommand{\bqa}{\begin{eqnarray}}
	\newcommand{\eqa}{\end{eqnarray}}

\makeatletter

\usepackage{babel}

\begin{document}
	
\title{Quantification of Wigner Negativity Remotely Generated via Einstein-Podolsky-Rosen Steering}
	
	\author{Yu~Xiang$^{\ddagger}$}
\address{State Key Laboratory for Mesoscopic Physics, School of Physics, Frontiers Science Center for Nano-optoelectronics, $\&$ Collaborative Innovation Center of Quantum Matter, Peking University, Beijing 100871, China}
\address{Collaborative Innovation Center of Extreme Optics, Shanxi University, Taiyuan, Shanxi 030006, China}
	\author{Shuheng~Liu$^{\ddagger}$}
	\address{State Key Laboratory for Mesoscopic Physics, School of Physics, Frontiers Science Center for Nano-optoelectronics, $\&$ Collaborative Innovation Center of Quantum Matter, Peking University, Beijing 100871, China}
	\author{Jiajie~Guo}
   \address{State Key Laboratory for Mesoscopic Physics, School of Physics, Frontiers Science Center for Nano-optoelectronics, $\&$ Collaborative Innovation Center of Quantum Matter, Peking University, Beijing 100871, China}
   	
	\author{Qihuang Gong}
\address{State Key Laboratory for Mesoscopic Physics, School of Physics, Frontiers Science Center for Nano-optoelectronics, $\&$ Collaborative Innovation Center of Quantum Matter, Peking University, Beijing 100871, China}
\address{Collaborative Innovation Center of Extreme Optics, Shanxi University, Taiyuan, Shanxi 030006, China}
\address{Peking University Yangtze Delta Institute of Optoelectronics, Nantong 226010, Jiangsu, China}

\author{Nicolas Treps}
\address{Laboratoire Kastler Brossel, Sorbonne Universit\'e, CNRS, ENS-Universit\'e PSL, Coll\`ege de France; 4 place Jussieu, F-75252 Paris, France}

	\author{Qiongyi~He}
	\email{qiongyihe@pku.edu.cn}
\address{State Key Laboratory for Mesoscopic Physics, School of Physics, Frontiers Science Center for Nano-optoelectronics, $\&$ Collaborative Innovation Center of Quantum Matter, Peking University, Beijing 100871, China}
\address{Collaborative Innovation Center of Extreme Optics, Shanxi University, Taiyuan, Shanxi 030006, China}
\address{Peking University Yangtze Delta Institute of Optoelectronics, Nantong 226010, Jiangsu, China}
   	\author{Mattia Walschaers}
\address{Laboratoire Kastler Brossel, Sorbonne Universit\'e, CNRS, ENS-Universit\'e PSL, Coll\`ege de France; 4 place Jussieu, F-75252 Paris, France}

\begin{abstract}
	Wigner negativity, as a well-known indicator of nonclassicality, plays an essential role in quantum computing and simulation using continuous-variable systems. Recently, it has been proven that Einstein-Podolsky-Rosen steering is a prerequisite to generate Wigner negativity between two remote modes. Motivated by the demand of real-world quantum network, here we investigate the shareability of generated Wigner negativity in the multipartite scenario from a quantitative perspective. By establishing a monogamy relation akin to the generalized Coffman-Kundu-Wootters inequality, we show that the amount of Wigner negativity cannot be freely distributed among different modes. Moreover, for photon subtraction---one of the main experimentally realized non-Gaussian operations---we provide a general method to quantify the remotely generated Wigner negativity. With this method,  we find that there is no direct quantitative relation between the Gaussian steerability and the amount of generated Wigner negativity. Our results pave the way for exploiting Wigner negativity as a valuable resource for numerous quantum information protocols based on non-Gaussian scenario.
\end{abstract}

\maketitle

Continuous-variable (CV) systems have attained impressive success in quantum information processing~\cite{cv}. As an important platform that has been widely studied, Gaussian systems and operations are extensively used in
quantum teleportation~\cite{tele}, quantum key distribution~\cite{qkd}, and quantum enhanced sensing~\cite{sensing1,sensing2}. These protocols come with the advantage of deterministically producing resource states and being analytically tractable due to the Gaussian properties of the states. However, non-Gaussian states and operations have irreplaceable advantages in several CV protocols~\cite{hybrid}, such as entanglement distillation~\cite{distill1,distill2}, error correction~\cite{error}, secure quantum communication~\cite{Nha_npj2019}, and the verification of Bell nonlocality~\cite{chenbell}. Considerable progresses in controllable generation of multimode non-Gaussian states have been made in recent experiments~\cite{npnongaussian,prxnongaussian}, which also provide support for the implementation of universal CV quantum computation in the long term~\cite{universalqc}. 

For some non-Gaussian states the Wigner function can reach negative values. This Wigner negativity has been seen as a necessary ingredient in CV quantum computating and simulation to outperform classical devices~\cite{Zhuangpra,Ferraropra,Eisertprl}. In the pursuit of networked quantum technologies it is crucial to develop efficient methods to produce Wigner negativity in the distant nodes. Recently, a scheme to remotely generate Wigner negativity was proposed through Einstein-Podolsky-Rosen (EPR) steering~\cite{WalschaersPRL,WalschaersPRXQ}---a particular type of quantum correlation where local measurements performed on one party can adjust (steer), instantaneously, the state of the other remote party~\cite{steering,Howard07PRL,ReidRMP}. Based on this kind of nonlocal effect, one can induce negativity in the steering mode by applying a set of appropriate operations on the steered mode. 

In consideration of the real-world quantum network in the multipartite scenario, it is a worthwhile objective to deeply explore the remote generation and distribution of Wigner negativity over many nodes in an entanglement-based network. As an intermediate type of quantum correlations between entanglement and Bell nonlocality, multipartite quantum steering~\cite{genuine13} has received extensive attention in recent developments of quantum information theory~\cite{cavalcanti17review,otreview}. It has been successfully implementated in CV optical network~\cite{ANUexp,prlSu,prlSu2020,yin}, photonic network~\cite{Spainexp,pan2015,pan2020}, and atomic ensembles~\cite{bec2}. Inspired by the shareability of EPR steering, known as monogamy~\cite{reidPRA,Yumonogamy,Kimmonogamy,GSmonogamy,Adesso16,SMmonogamy,SMmonogamyexp}, it is interesting to explore how can the remotely generated Wigner negativity be distributed over different modes? Is there any monogamy relations imposing quantitative constraints on that negativity? And does stronger steerability generate more negativity? 

\begin{figure}[tb]
	\begin{center}
		\includegraphics[width=70mm]{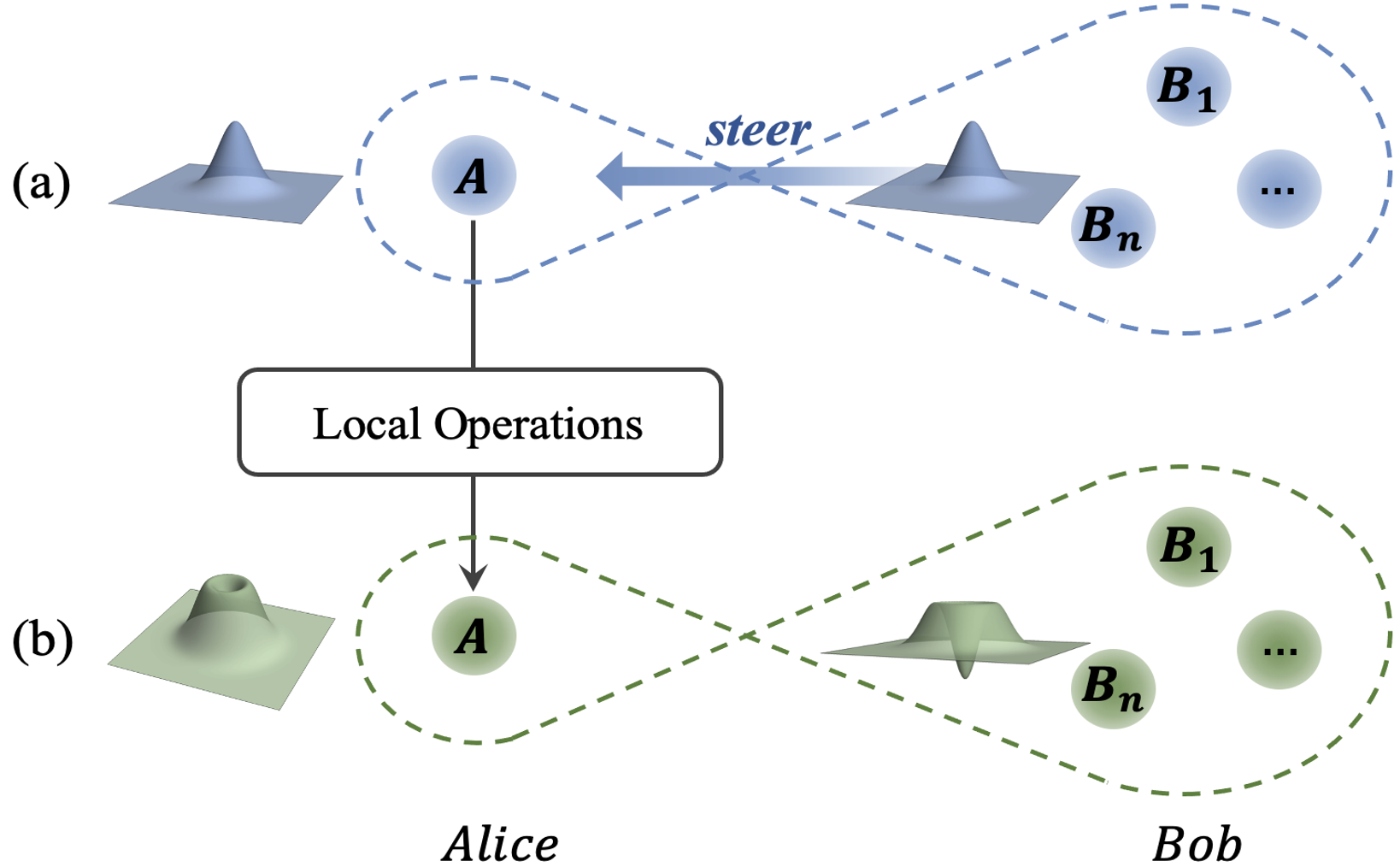}
	\end{center}
	\caption{Scheme of the remote generation of Wigner negativity through EPR steering in a multipartite scenario. (a) The initial Gaussian steerable system; (b) After some appropriate local operations on the steered mode hold by Alice, the steering subsystem hold by Bob becomes non-Gaussian with Wigner negativity.}
	\label{fig1}
\end{figure}

In this work we present a quantitative investigation of Wigner negativity that is remotely created via multipartite EPR steering, in which non-Gaussian operations performed on one steered node of quantum network produce Wigner negativity in different distant nodes, as shown in Fig.~\ref{fig1}. We first investigate to what extent Wigner negativity can be shared by establishing a monogamy relation. This constraints the degree of distributed negativity akin to the Coffman-Kundu-Wootters (CKW) monogamy inequality for steerability~\cite{Yumonogamy}. Then we focus on photon subtraction, a commonly used non-Gaussian operation, and find a general measure for the amount of induced Wigner negativity in the steering modes. This allows us to disprove the conjecture that stronger steerability creates more negativity. Specifically, considering the major experimentally realized CV EPR resources such as the two-mode EPR states with phase-insensitive losses and the two-mode squeezed thermal states, we show that the procedure for remote generation of Wigner negativity can be significantly simplified when the covariance matrix (CM) is transformed into its standard form. This provides a more insightful approach than previously proposed procedures~\cite{WalschaersPRL}, and makes such states particularly resourceful for remotely producing Wigner negativity. 

\textit{Multimode CV systems.--} We begin by briefly introducing the theoretical framework of multimode CV quantum optics. The noninteracting quantized electromagnetic field can be treated as a number $N$ of optical modes that behave as quantum harmonic oscillators with different frequencies described by $\hat{H}=\sum_{k=1}^{N}2\omega_k(\hat{a}_k^{\dagger}\hat{a}_k+\frac12)$.
Here, $\hat{a}_k$ and $\hat{a}_k^{\dagger}$ are the annihilation and creation operators of a photon in mode $k$, satisfying the bosonic commutation relation $[\hat{a}_k,\hat{a}_{k^\prime}^{\dagger}]=\delta_{kk^\prime}$. The corresponding quadrature phase operators for each mode are defined as $\hat{x}_k=\hat{a}_k+\hat{a}_{k}^{\dagger}$ and $\hat{p}_k=(\hat{a}_k-\hat{a}_{k}^{\dagger})/i$. Collecting the quadrature operators for all the modes into a vector $\hat{\boldsymbol{\xi}}\equiv (\hat{x}_{1},\hat{p}_{1},...,%
\hat{x}_{N},\hat{p}_{N})^\top$, the CM $\sigma$ is given with elements $\sigma _{ij}=\langle \hat{\xi}_{i}\hat{\xi}_{j}+\hat{\xi}_{j}
\hat{\xi}_{i}\rangle /2-\langle \hat{\xi}_{i}\rangle \langle \hat{\xi}%
_{j}\rangle$. If the system is prepared in a Gaussian state, the properties can be completely determined by its CM. Otherwise, the first and second-order statistical moments are not enough to characterize the non-Gaussian system, and we must resort to a more complete description. Here we choose the Wigner function as a preferred phase space representation for an arbitrary state with density matrix $\hat \rho$,
\begin{equation}
W({\boldsymbol{\xi}})=\int_{\mathbb{R}^{2 N}} \frac{d^{2 N} \boldsymbol{\alpha}}{(2 \pi)^{2 N}} \exp \left(-i {\boldsymbol{\xi}}^{\top} \boldsymbol{\Omega} \boldsymbol{\alpha}\right) \chi(\boldsymbol{\alpha}),
\end{equation}
where $\boldsymbol{\Omega}=\bigoplus_{1}^{N}\left(\begin{array}{cc}0 & 1 \\ -1 & 0\end{array}\right)$ is the symplectic form and the Wigner characteristic function $\chi(\boldsymbol{\alpha})=\text{Tr}[\hat{\rho}\exp(i\hat{\boldsymbol{\xi}}^{\top}\boldsymbol{\Omega}\boldsymbol{\alpha})]$. A particular attribute of non-Gaussian states is the possibility for this Wigner function to attain negative values that can be quantified as  $\mathcal{N}\equiv\int |W({\boldsymbol{\xi}})| d {\boldsymbol{\xi}}-1$ \cite{Kenfack}. 

\textit{Remote generation of Wigner negativity through multipartite EPR steering.--} In order to effectively generate and distribute Wigner negativity, an indirect scheme was proposed based on EPR steering~\cite{WalschaersPRL}.  
In a two-mode Gaussian system, when there exists steering from Bob to Alice, then an appropriate local Gaussian transformation together with photon subtraction on the steered mode $A$ can remotely generate Wigner negativity in the steering mode $B$, i.e. $\mathcal{N}_B>0$. The bipartite Gaussian steerability can be quantified by the parameter $\mathcal{G}^{B\rightarrow A}=\max\{0, \frac12\ln\frac{\text{Det~}\sigma_{B}}{\text{Det~}\sigma_{AB}}\}$, where $\sigma_{B}$ and $\sigma_{AB}$ denote the CM for mode $B$, and the group $(AB)$, respectively~\cite{Adesso15}. This formalism was developed for arbitrary conditional operations on an arbitrary number of modes, showing that EPR steering is still necessary to prepare a Wigner-negative state in the steering modes~\cite{WalschaersPRXQ}. The remotely generated Wigner negativity was not quantified, nor are its multimode properties such as the shareability of the negativity among steering modes understood. Especially, since EPR steering is a prerequisite for remote preparation of Wigner negativity, one may intuitively expect that stronger steerability in the initial Gaussian states creates more Wigner negativity. With our quantitative investigation, we show that conjecture is not the case. 

\textit{Monogamy of remotely generated Wigner negativity.--} First, we study the multimode character of the remotely generated Wigner negativity by deriving constraints on the distribution of this negativity among various modes in the steering party $(B_1B_2\ldots B_n)$ for a $(1+n)$-mode Gaussian state $\sigma_{AB_1B_2\ldots B_n}$. As a fundamental property of EPR steering, the CKW-type monogamy relation reveals that the sum of Gaussian steerability between any two modes cannot exceed their intergroup steerability, i.e., $\mathcal{G}^{B_1B_2\ldots B_n\rightarrow A}\geq \sum_{i=1}^{n}\mathcal{G}^{B_i\rightarrow A}$, which bounds the key rate in quantum secret sharing~\cite{Yumonogamy}. In analogy with the steering constraint, we establish a monogamy relation for the amount of the generated Wigner negativity,
\begin{equation}
\mathcal{N}_{B_1B_2\ldots B_n}(\mathcal{L}_{A|B_1B_2\ldots B_n})\geq \sum_{i=1}^{n}\mathcal{N}_{B_i}(\mathcal{L}_{A|B_i}),
\label{mono}
\end{equation}
where $\mathcal{N}_{B_j \ldots B_k}$ denotes the Wigner negativity created in the set of modes $(B_j \ldots B_k)$ by performing some appropriate operations $\mathcal{L}_{A|B_j \ldots B_k}$ on subsystem $A$. Note that generating negativities in different set of modes requires different optimal operations on the steered mode $A$. For instance, inducing Wigner negativity in the steering mode $B_j$, or $B_k$, or their joint $(B_jB_k)$, requires different local Gaussian transformations prior to a non-Gaussian operation.
\begin{figure}[t]
	\begin{center}
		\includegraphics[width=87mm]{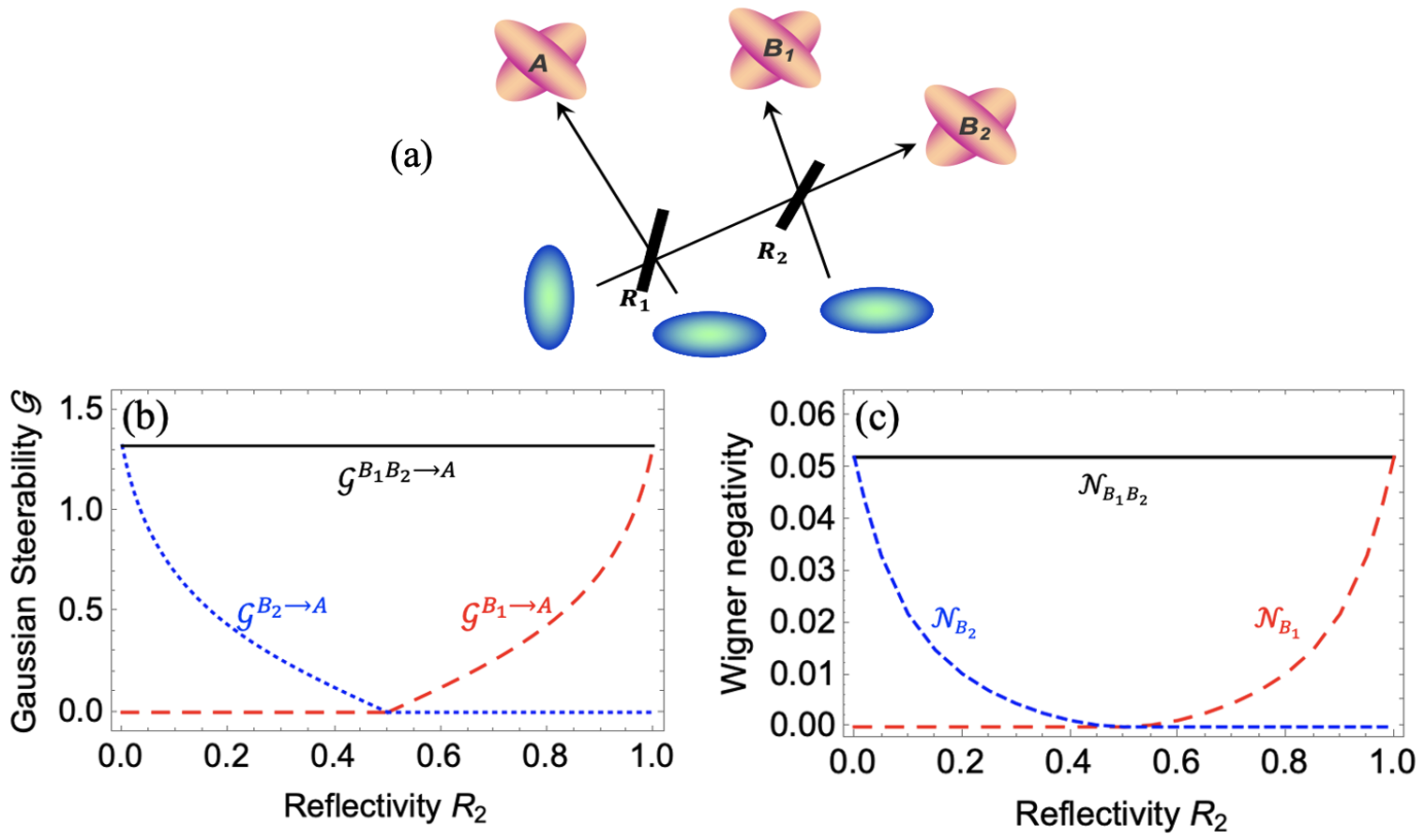}
	\end{center}
	\caption{(a) Scheme of a pure three-mode Gaussian state generated by the linear optical network. (b) Fixing $R_1:(1-R_1)=50:50$ and input squeezing levels $r=1$ (corresponding to $-8.7{\rm dB}$ quadrature noise), the initial Gaussian steerability changes with a variable $R_2$. (c) After a single-photon subtraction on mode $A$, the remotely generated Wigner negativity of mode $B$, mode $C$ and the group $(BC)$, respectively.}
	\label{fig2}
\end{figure}

To prove the above inequality we use the fact that $\mathcal{N}_{B_j}(\mathcal{L}_{A|B_j})>0$ and $\mathcal{N}_{B_k}(\mathcal{L}_{A|B_k})>0$ ($j\neq k$) cannot be true simultaneously, which is a consequence of another type of Gaussian steering monogamy relation: modes $B_j$ and $B_k$ cannot simultaneously steer mode $A$ under Gaussian measurements~\cite{reidPRA,Kimmonogamy}. Assuming that mode $B_1$ can steer mode $A$, negativity can be generated only in the Wigner function of mode $B_1$ under Alice's local operation $\mathcal{L}_{A|B_1}$. We then just need to prove that $\mathcal{N}_{B_1B_2\ldots B_n}(\mathcal{L}_{A|B_1B_2\ldots B_n}) \geq \mathcal{N}_{B_1}(\mathcal{L}_{A|B_1})$. The detailed proof is deferred to the Supplemental Material~\cite{Supp}.  An example is given in Fig.~\ref{fig2}. Here the local non-Gaussian operation we chose for Alice is single-photon subtraction $\mathcal{S}$, which can be effectively realized in experiments~\cite{npnongaussian,scisubtraction}. In this system, a pure three-mode entangled Gaussian state can be generated by the linear optical network with three squeezed inputs, as illustrated in Fig.~\ref{fig2}(a). When the first beam splitter is fixed at $R_1:(1-R_1)=50:50$ and the second beam splitter is adjustable, the steerability distributed among three modes in the initial Gaussian state and the corresponding Wigner negativities remotely created through photon subtraction on mode $A$ are shown in Figs.~\ref{fig2}(b) and (c), respectively. It is clear that the two-mode and three-mode Gaussian steering $\mathcal{G}^{B_1\rightarrow A}$, $\mathcal{G}^{B_2\rightarrow A}$ and $\mathcal{G}^{B_1B_2\rightarrow A}$ are necessary to induce Wigner negativity for individual mode $B_1$, $B_2$, and for the group $(B_1B_2)$, respectively. In addition, we observe that, even though $\mathcal{N}_{B_1}(\mathcal{S}_{A|B_1})>0$ and $\mathcal{N}_{B_2}(\mathcal{S}_{A|B_2})>0$ cannot be satisfied at the same time, the joint Wigner negativity created on the group $(B_1B_2)$ is significantly higher than the negativity in either individual mode.

\textit{Quantification of the generated Wigner negativity.--} Since remotely generated Wigner negativity is both enabled and constrained by Gaussian steering, one may intuitively expect that stronger steerability induces more Wigner negativity. To show that --unexpectedly-- this is not the case, we quantify the amount of Wigner negativity in the steering modes, via the purities of initial Gaussian states. We present an explicit study for some experimentally prominent two-mode Gaussian states and show that purity, rather than steerability, governs the amount of Wigner negativity that can be created.

It is well known that any two-mode Gaussian state can be transformed into a \textit{standard form}~\cite{standard} through local linear unitary Bogoliubov operations (LLUBOs), so that the CM $\sigma_{AB}$ reads
\begin{equation}
\sigma_{AB, sf}=\left(
\begin{array}{cc}
\sigma_{A}  & \gamma_{AB}\\
\gamma_{AB}^{\top} &\sigma_{B}\\
\end{array}
\right)=\left(
\begin{array}{cccc}
a & 0 & c_1 & 0\\
0 & a & 0 & c_2\\
c_1 & 0 & b & 0\\
0 & c_2 & 0 & b\\
\end{array}
\right)\label{sf2}
\end{equation}
with $a,~b\geq1$ and $ab-c_{1(2)}^2\geq0$. The two local purities $\mu_{A(B)}\equiv 1/\sqrt{\text{Det~}\sigma_{A(B)}}=1/a(b)$ and the global purity $\mu_{AB}\equiv 1/\sqrt{\text{Det~}\sigma_{AB}}=1/[(ab-c_1^2)(ab-c_2^2)]$, are invariant under LLUBOs~\cite{pra73}. Furthermore, because LLUBOs are local, the standard form $\sigma_{AB,sf}$ manifests equal quantity of Gaussian steerability possessed in the initial states.

We now restrict our scope to two-mode Gaussian states $c_1=-c_2=c$, which include states generated by parametric down-conversion. By focusing on the experimentally relevant case where a photon subtracted from the steered mode, we derive that the amount of remotely generated Wigner negativity $\mathcal{N}_B$ is determined by the purities of initial Gaussian state $\mu_A$, $\mu_B$ and $\mu_{AB}$: 
\begin{equation}
\mathcal{N}_{B}(\mathcal{S}_{A|B})=2\left[ \frac{e^{\frac{\mu_A\mu_B-\mu_{AB}\mu_A}{\mu_{AB}-\mu_A\mu_B}}(\mu_A\mu_B-\mu_{AB})}{\mu_{AB}(\mu_A-1)}-1\right].\label{purity-WN}
\end{equation}
Exchanging $\mu_A \leftrightarrow \mu_B$ we can obtain the result for the other direction $\mathcal{N}_A(\mathcal{S}_{B|A})$. The derivation of the above relation is detailed in the Supplemental Material \cite{Supp}. EPR steering and remotely induced Wigner negativity are both determined by the local and global purities, but their dependence on these purities is very different. EPR steering provides a necessary bridge to induce Wigner negativity, but it is insufficient to unambiguously quantify the created Wigner negativity. 

\begin{figure}[!tbp]
	\begin{center}
		\includegraphics[width=87mm]{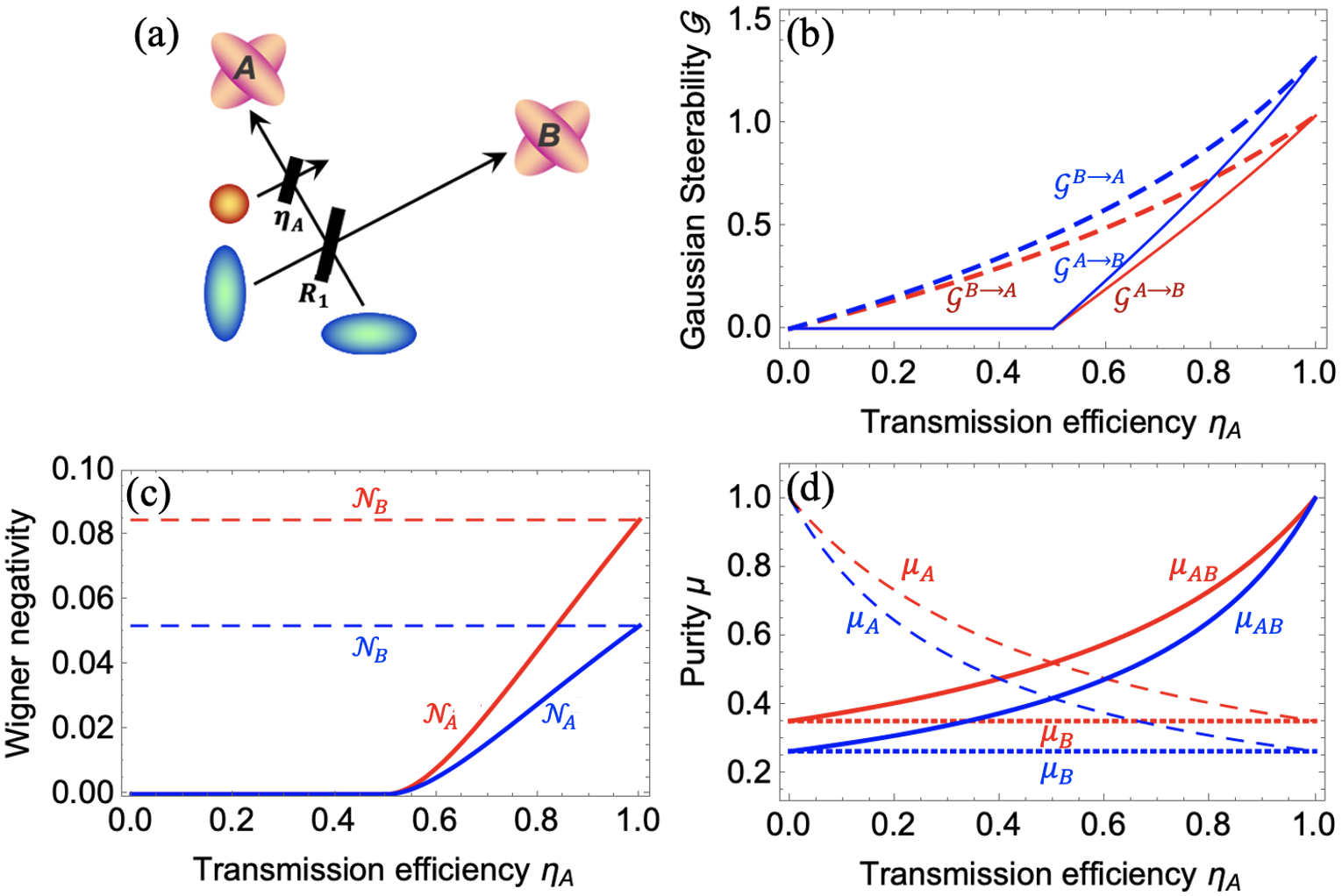}
	\end{center}
	\caption{(a) Scheme of a two-mode squeezed vacuum state with one lossy channel on mode $A$, where $R_{1}$ is a balanced beam splitter. (b) The initial asymmetric Gaussian steerability with different squeezing levels $r=1$ (blue) and $r=0.85$ (red), corresponding to a quadrature noise reduction of $-8.7{\rm dB}$ and $-7.4 {\rm dB}$, respectively. Loss has a more significant effect on the steering mode. (c) After the single-photon subtraction on the steered mode, the amount of induced Wigner negativity on the steering mode corresponding to Gaussian steerability given in (b). (d) The local and global purities of the initial Gaussian states with different squeezing levels.}
	\label{fig3}
\end{figure}

We explicitly show this point in Fig.~\ref{fig3}, where we consider a two-mode EPR state with one lossy channel on mode $A$, which is often used to demonstrate one-way steering~\cite{ANUexp,prlSu,OneWayNatPhot}. This state is already in the standard form (\ref{sf2}), with $a=\eta_A(\cosh 2r-1)+1$, $b=\cosh 2r$, and $c_1=-c_2=\sqrt{\eta_A}\sinh{2r}$, where $r$ is the squeezing parameter and $\eta_A$ is the transmission efficiency. The asymmetric Gaussian steerabilities in two directions are indicated in Fig.~\ref{fig3}(b). We find that the Gaussian steerability $\mathcal{G}^{A\rightarrow B}>0$ with a threshold level of efficiency $\eta_A>0.5$ and becomes stronger with a higher squeezing level. Thus, by performing a single-photon subtraction on the steered mode $B$, the Wigner negativity $\mathcal{N}_{A}(\mathcal{S}_{B|A})$ can be generated when $\eta_A>0.5$ as well and becomes larger with increasing efficiency $\eta_A$, as shown in Fig.~\ref{fig3}(c). The Gaussian steerability in the other direction $\mathcal{G}^{B\rightarrow A}>0$ happens for any $\eta_A>0$ and enhances with a higher squeezing level. Interestingly, for this direction, by performing a single-photon subtraction on the steered mode $A$, the generated negativity $\mathcal{N}_{B}(\mathcal{S}_{A|B})$ does not vary with $\eta_A$. From Eq.~(\ref{purity-WN}), we can obtain that $\mathcal{N}_B(\mathcal{S}_{A|B})=2e^{-\mu_B/(\mu_B+1)}(\mu_B+1)-2$, which is solely determined by the local purity $\mu_B$. It is unchanged since no loss is considered in mode $B$. While the value of $\mathcal{N}_A(\mathcal{S}_{B|A})$ depends on $\mu_{A,~B,~AB}$ and thus varies with $\eta_A$, as shown in Fig.~\ref{fig3}(d). This highlights the asymmetry of the induced Wigner negativity and suggests a way to remotely generate negativity that is robust to channel loss. Although the states produced by a higher squeezing level (blue lines) possess stronger Gaussian steerability, their purities are more sensitive to the loss~\cite{purityG}. Therefore, the created Wigner negativity in the steering mode is larger when the initial squeezing level is lower.

\begin{figure}[b]
	\begin{center}
		\includegraphics[width=70mm]{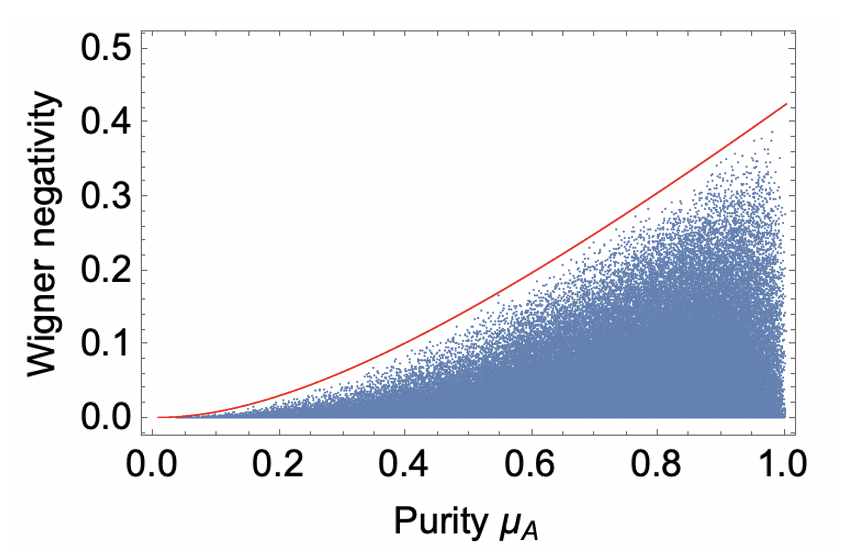}
	\end{center}
	\caption{The induced Wigner negativity $\mathcal{N}_B(\mathcal{S}_{A|B})$ for $\sim250000$ randomly mixed two-mode Gaussian states (dots) are bounded by Eq.~(\ref{pure}) (red curve).}
	\label{fig4}
\end{figure}

Moreover, for any globally pure state, i.e., $\mu_{AB}=1$, with $\mu_A=\mu_B=\mu$, Eq.~(\ref{purity-WN}) is simplified to be
\begin{equation}
\mathcal{N}_B(\mathcal{S}_{A|B})=\mathcal{N}_A(\mathcal{S}_{B|A})=2\left[e^{-\frac{\mu}{1+\mu}}(1+\mu)-1\right].
\label{pure}
\end{equation}
We show that Eq.~\ref{pure} bounds the remotely created Wigner negativity for arbitrary mixed two-mode states. In Fig.~\ref{fig4}, we plot 250000 dots which present the Wigner negativities of randomly generated by photon subtraction from two-mode Gaussian states. All dots are located below the red curve described by Eq.~\ref{pure}. Furthermore, the CM for an arbitrary three-mode pure state $\sigma_{A-(B_1B_2)}^{\text{pure}}$ with respect to $A-(B_1B_2)$ splitting is locally equivalent to 
\begin{equation}
\sigma_{A-(B_1B_2),sf}^{\text{pure}}=\left(
\begin{array}{cccccc}
a & 0 & c_1 & 0 & 0 & 0\\
0 & a & 0 & -c_1& 0 & 0\\
c_1 & 0 & a & 0 & 0 & 0\\
0 & -c_1 & 0 & a& 0 & 0\\
0 & 0 & 0 & 0 & 1 & 0\\
0 & 0 & 0 & 0 & 0 & 1
\end{array}
\right)\label{sfabc}
\end{equation} 
through Williamson decomposition~\cite{jpaGA}, where $c_1=\sqrt{a^2-1}$. We can then see that the system is expressed by a product of a two-mode squeezed vacuum state tensor an uncorrelated vacuum mode, such that steering property between mode $A$ and modes $(B_1B_2)$ can be unitarily reduced to the two-mode case. Thus, the generated Wigner negativity of the joint group $(B_1B_2)$ is equivalently indicated by the red curve in Fig.~\ref{fig4}, which is always higher than the negativity induced in the individual mode. Besides the examples discussed above, we also analyze the asymmetric Gaussian steerability and the properties of induced Wigner negativity for another important CV EPR resource--two-mode squeezed thermal states~\cite{He15}, which are detailed in the Supplemental Material~\cite{Supp}.

Finally, we stress that when the CM of a bipartite system is transformed in the standard form and satisfies $c_1=-c_2=c$, photon subtraction on mode $A$ can always generate Wigner negativity in mode $B$ as long as $\mathcal{G}^{B\rightarrow A}>0$. The proof is deferred in the Supplemental Material~\cite{Supp}. Thus producing EPR resource with CM in a \textit{standard form} significantly simplifies the procedure for remote generation of Wigner negativity and makes the resulting non-Gaussian state readily available for further applications. This complements the results of Ref.~\cite{WalschaersPRL}, where it was shown that an additional local Gaussian transformation prior to photon subtraction is necessary to make EPR steering sufficient for remotely generating Wigner negativity. This Gaussian transformation requires inline squeezing and is experimentally challenging. Our results can be used as a recipe to prepare resourceful Gaussian states for the remote generation of Wigner negativity without the need for inline squeezing.

\textit{Conclusion.}-- We develop the scheme for remote generation of Wigner negativity through EPR steering to multimode scenario, and show the presence of constraints for distributing Wigner negativity over different modes. So far, multipartite steering has been demonstrated in various Gaussian systems, e.g., linear optical networks~\cite{ANUexp,prlSu}, quantum frequency comb~\cite{yin}, and Bose-Einstein condensates~\cite{bec2}. These experimental developments lay a favorable foundation to implement remote generation of multipartite non-Gaussian states through photon subtraction or other appropriate operations. Furthermore, we present an intuitive and computable quantification of the generated Wigner negativity for bipartite system in terms of the local and global purities of initial Gaussian state. Our results deepen the understanding of Wigner negativity as a resource and provide an important framework of non-Gaussian quantum information theory.  

Our work also triggers several new questions to stimulate further research. For instance, as Gaussian steerability $\mathcal{G}^{A\rightarrow B_1}>0$ and $\mathcal{G}^{A\rightarrow B_2}>0$ can happen simultaneously, then by performing a single-photon subtraction on each mode $B_1$, $B_2$, can we achieve more significant increase of the negativity in mode $A$? In addition, for this direction the Gaussian steerability still follows the CKW-type monogamy constraint, however, this constraint does not hold any more for the generated negativity. We have observed a violation in a pure three-mode state~\cite{Supp}, i.e., $\mathcal{N}_{A}(\mathcal{L}_{B_1B_2|A})<\mathcal{N}_{A}(\mathcal{L}_{B_1|A})+\mathcal{N}_{A}(\mathcal{L}_{B_2|A})$. Moreover, after non-Gaussian operations on the steered mode, the resulting system cannot be fully captured by the second-order correlations given in CM. To this day, relatively little is known about the characteristics of non-Gaussian steering~\cite{Yunon}. 

\begin{acknowledgments}
	This work is supported by the National Natural Science Foundation of China (Grants No. 11975026, No. 61675007, and No. 12004011), the National Key R$\&$D Program of China (Grants No. 2016YFA0301302, No. 2018YFB1107205, and 2019YFA0308702). Q.H. also thanks partial support from Beijing Natural Science Foundation (Grant No. Z190005) and the Key R$\&$D Program of Guangdong Province (Grant No. 2018B030329001). N.T. and M.W. received funding from the European Union’s Horizon 2020 research and innovation programme under grant agreement No 899587.\\
	\newline
	$^{\ddagger}$Y. Xiang and S. H. Liu contributed equally to this work.

\end{acknowledgments}

\appendix*
\setcounter{figure}{0}
\renewcommand\thefigure{A\arabic{figure}}
\setcounter{table}{0}
\renewcommand\thetable{A\arabic{table}}

\section*{Appendix A: Monogamy of Wigner negativity remotely created through EPR steering}
Monogamy is known as a fundamental property of multipartite quantum correlations, which has profound applications in quantum communication. In this Section, we prove that the remotely created Wigner negativity cannot be freely shared among different parties by establishing a monogamy relation in direct analogy with the Coffman-Kundu-Wootters (CKW) inequality for EPR steering~\cite{Yumonogamy}, as given by Eq.~(2) in the main text.

Without loss of generality, let us focus on a tripartite scenario, in which the steering party $B$ contains two modes $B_1$ and $B_2$, such that the monogamy inequality takes the simpler form, 
\begin{equation}
\label{eMonogamyBC}
\mathcal{N}_{B_1 B_2}(\mathcal{L}_{A|B_1B_2})\geqslant\mathcal{N}_{B_1}(\mathcal{L}_{A|B_1})+\mathcal{N}_{B_2}(\mathcal{L}_{A|B_2}),
\end{equation}
when the subsystem $A$ can be steered by the group $(B_1 B_2)$. By taking $A$ being a single mode and $B_1, B_2$ comprising an arbitrary number of modes, one can then apply iteratively above inequality to obtain the corresponding general multipartite inequality (2) in the main text.

To do so, recall the Gaussian steering monogamy relations from Refs.~\cite{reidPRA,Kimmonogamy} that it is impossible for modes $B_1$ and $B_2$ to simultaneously steer the mode $A$ under Gaussian measurements, i.e.,  $\mathcal{N}_{B_1}(\mathcal{L}_{A|B_1})>0$ and $\mathcal{N}_{B_2}(\mathcal{L}_{A|B_2})>0$ cannot be true simultaneously. Granted that $B_1$ can steer mode $A$, the monogamy relation (\ref{eMonogamyBC}) reduces to $\mathcal{N}_{B_1 B_2}(\mathcal{L}_{A|B_1B_2})\geqslant\mathcal{N}_{B_1}(\mathcal{L}_{A|B_1})$ (or the analogous expression with swapped $B_1\leftrightarrow B_2$). Because the Wigner negativity is nonincreasing under partial trace on the steering party $(B_1 B_2)$~\cite{Ferraropra}, we have 
\begin{equation}
\label{eBCTrace}
\mathcal{N}_{B_1 B_2}(\mathcal{L}_{A|B_1})\geqslant \mathcal{N}_{B_1}(\mathcal{L}_{A|B_1}).
\end{equation}
Note that the left and right sides of above inequality are given by the same resulting state $\rho_{B_1 B_2}$ after a local operation $\mathcal{L}_{A|B_1}$ on the steered mode $A$. The proof is straightforward~\cite{Ferraropra}, as given by
\begin{equation}
\begin{aligned}
\mathcal{N}_{B_1}(\mathcal{L}_{A|B_1})&=\mathcal{N}[\operatorname{Tr}_{B_2}\rho_{B_1 B_2}]\\
&=\int d \boldsymbol{r}_{B_1}\left| W[\operatorname{Tr}_{B_2}[\rho_{B_1 B_2}]](\boldsymbol{r}_{B_1})\right|-1\\
&=\int d \boldsymbol{r}_{B_1}\left| \int d \boldsymbol{r}_{B_2} W[\rho_{B_1 B_2}](\boldsymbol{r}_{B_1},\boldsymbol{r}_{B_2})\right|-1\\
&\leqslant \int d \boldsymbol{r}_{B_1} \int d \boldsymbol{r}_{B_2} \left| W[\rho_{B_1 B_2}](\boldsymbol{r}_{B_1},\boldsymbol{r}_{B_2})\right|-1\\
&=\mathcal{N}[\rho_{B_1 B_2}]=\mathcal{N}_{B_1 B_2}(\mathcal{L}_{A|B_1}),
\end{aligned}
\end{equation}
where $W[\rho_{B_1B_2}](\boldsymbol{r})$ represents the Wigner distribution of state $\rho_{B_1 B_2}$. Afterwards it is promptly verified that there always exists an optimized local operation $\mathcal{L}_{A|B_1B_2}$ can generated the largest Wigner negativity $\mathcal{N}_{B_1 B_2}(\mathcal{L}_{A|B_1B_2})$ to make $\mathcal{N}_{B_1 B_2}(\mathcal{L}_{A|B_1B_2}) \geq \mathcal{N}_{B_1B_2}(\mathcal{L}_{A|B_1}) \geq \mathcal{N}_{B_1}(\mathcal{L}_{A|B_1})$.

\section*{Appendix B: Quantifying the remotely created Wigner negativity}
It is of particular interest to us is whether stronger steerability in the initial Gaussian states induces more Wigner negativity, as it is enabled and constrained by Gaussian steering. To answer this, we need first quantify the amount of Wigner negativity. In this part, we aim to derive the qualitative measure of Wigner negativity Eq.~(4) in the main text by focusing on the experimentally relevant case where a photon subtracted from the steered mode in two-mode Gaussian states $c_1=-c_2=c$, which includes the major experimentally realized CV EPR states such as the two-mode squeezed vacumm states with phase-insensitive losses and the two-mode squeezed thermal states (STS).  

Let us recall that, any two-mode Gaussian state can be transformed through LLUBOs to the standard form
\begin{equation}
\sigma_{AB, sf}=\left(
\begin{array}{cc}
\sigma_{A}  & \gamma_{AB}\\
\gamma_{AB}^{\top} &\sigma_{B}\\
\end{array}
\right)=\left(
\begin{array}{cccc}
a & 0 & c_1 & 0\\
0 & a & 0 & c_2\\
c_1 & 0 & b & 0\\
0 & c_2 & 0 & b\\
\end{array}
\right).
\end{equation}
Using the formula derived from Ref.~\cite{WalschaersPRL}, we obtain the reduced Wigner function of the steering mode $B$ after the local Gaussian transformation $R_{A|B}$ combined with a single-photon subtraction applied on the steered mode $A$,
\begin{equation}
\begin{aligned}
W_{B}\left(\beta_{B}\right)&=\frac{\exp 
	\left\{-\frac{1}{2}\left(\beta_{B}, \sigma_{B}^{-1}\beta_{B}\right)\right\}}{2 \pi \sqrt{\operatorname{Det} 
		\sigma_{B}}\left[\operatorname{Tr}\left(R_{A|B}^{\top} \sigma_{A} R_{A|B}\right)-2\right]}\times\\
&~~\left[\beta_{B}^{\top} {\sigma_{B}^{-1}}^{\top} \gamma_{AB}^{\top} R_{A|B} R_{A|B}^{\top} \gamma_{AB} \sigma_{B}^{-1}\beta_{B}+\operatorname{Tr}(R_{A|B}^{\top} V_{A \mid  B} R_{A|B})-2\right]\\
&=\frac{\exp \left\{-\frac{1}{2}\left(\beta_{B}, \sigma_{B}^{-1}\beta_{B}\right)\right\}}{2 \pi \sqrt{\operatorname{Det} \sigma_{B}}\left[\operatorname{Tr}\left(\nu \sigma_{A} V_{A \mid  B}^{-1}\right)-2\right]}\times\\
&~~\left(\nu \beta_{B}^{\top} {\sigma_{B}^{-1}}^{\top} \gamma_{AB}^{\top} V_{A \mid  B}^{-1} \gamma_{AB} \sigma_{B}^{-1}\beta_{B}+2\nu-2\right),
\end{aligned}
\end{equation}
where $\vec{\beta}_{B}$ is the coordinate in a multimode phase spaces of subsystem $B$, $V_{A \mid  B}=\sigma_{A}-\gamma_{AB} \sigma_{B}^{-1} \gamma_{AB}^{\top}$ is the Schur complement of $\sigma_B$ and $\nu$ is the corresponding symplectic eigenvalue. The Schur complement $V_{A \mid  B}$ can be decomposed through Williamson decomposition via $V_{A \mid  B}=\nu S_{A|B}^{\top}S_{A|B}$, where $S_{A|B}$ is the corresponding symplectic matrix and a local Gaussian transformation $R_{A|B}=S_{A|B}^{-1}$.
When it comes to our particular interest subclass $c_1=-c_2=c$, it is easy to find out that the Schur complement is a multiple of identity matrix so that there is no need to perform an additional local Gaussian operation. Then we get
\begin{equation}
W_{B}\left(\rho\right)=\frac{e^{-\frac{\rho ^2}{2 b}} \left(2 b^2 (a-1)-2 b c^2+c^2 \rho ^2\right)}{4 \pi  b^3 (a-1)},
\end{equation}
which is circular symmetric. It is straightforward to calculate Wigner negativity using integral,
\begin{equation}
\label{eWignerNegativityabc}
\mathcal{N}_{B}(\mathcal{S}_{A|B})=\frac{2 c^2 e^{\frac{b (a-1)}{c^2}-1}}{b (a-1)}-2.
\end{equation}
By expressing $\mathcal{N}_{B}(\mathcal{S}_{A|B})$ in terms of purity $\mu_{AB}=1/(a b-c^2),~\mu_{A}=1/a,~\mu_{B}=1/b$, Eq.~(\ref{eWignerNegativityabc}) becomes
\begin{equation}
\label{WN}
\mathcal{N}_{B}(\mathcal{S}_{A|B})=2\left[\frac{ e^{\frac{ \mu_{A}\mu_{B}-\mu_{AB}\mu_{A}}{\mu_{AB}-\mu_A \mu_{B}}}(\mu_{A} \mu_{B}-\mu_{AB})}{ \mu_{AB}(\mu_{A}-1)}-1\right].
\end{equation}
\begin{figure}[]
	\begin{center}
		\includegraphics[width=78mm]{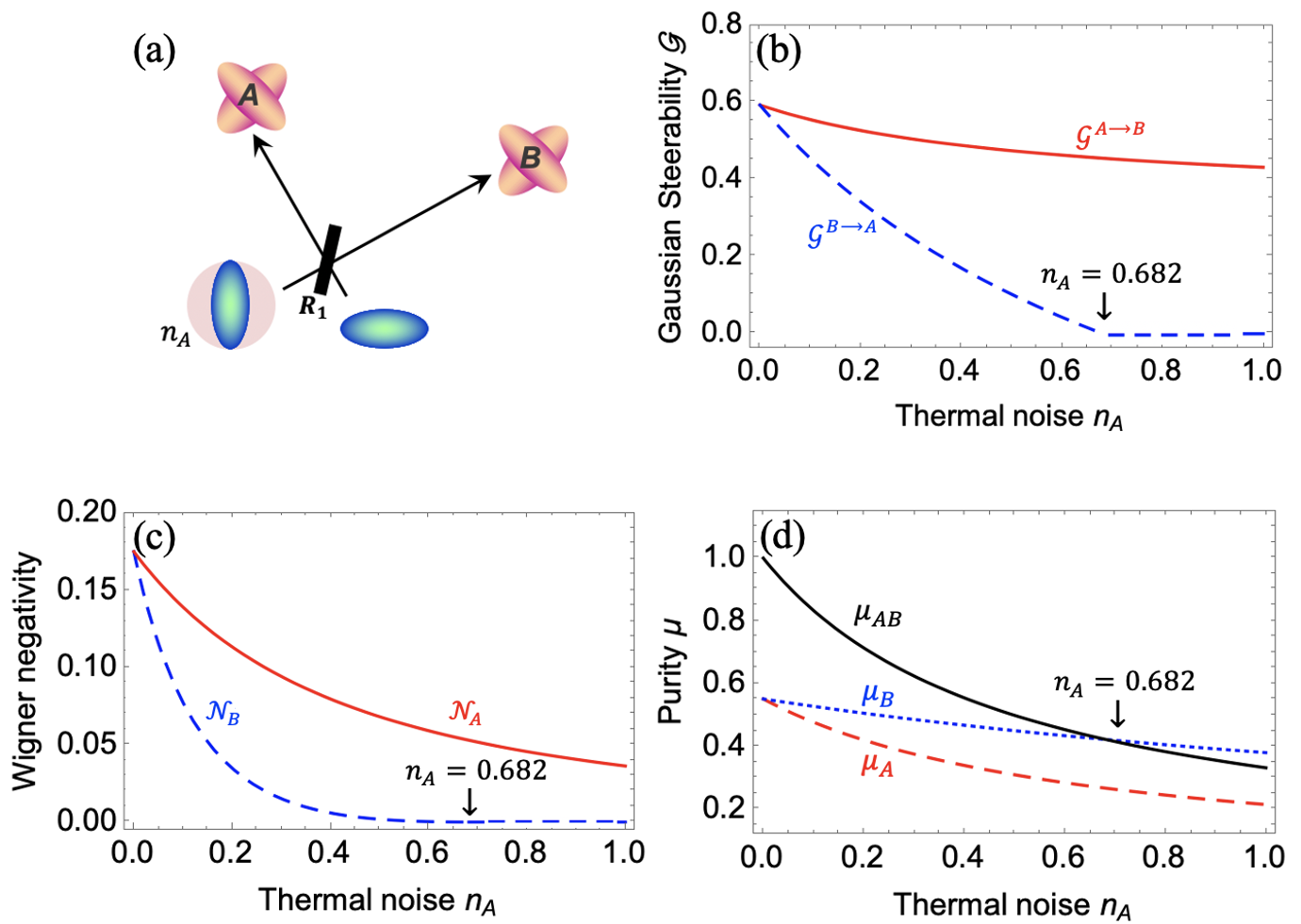}
	\end{center}
	\caption{(a) Scheme of a two-mode squeezed thermal state with asymmetric thermal noise $n_A$ and $n_B=0$. (b) The initial asymmetric Gaussian steerability with fixed squeezing level of $r=0.6$ (corresponding to $5.2{\rm dB}$ quadrature noise), where thermal noise has a more significant effect on the steered mode. (c) Corresponding to (b), after the single-photon subtraction on one side, the remotely generated Wigner negativity of the other side. (d) The local and global purities of the initial Gaussian states.}
	\label{figS1}
\end{figure}

Figure 3 in the main text showed the case of two-mode squeezed vacuum state with one lossy channel on one mode. Here, we particularly show the results for the two-mode squeezed thermal states. The CM elements of these states are $a=(n_A+n_B+1)\cosh(2r)+(n_A-n_B)$, $b=(n_A+n_B+1)\cosh(2r)-(n_A-n_B)$, $c_1=-c_2=(n_A+n_B+1)\sinh(2r)$, where $n_{A}$, $n_{B}$ are the average number of thermal photons for each subsystem~\cite{He15}. We set the thermal noise only on mode $A$ and leave $n_{B}=0$, as illustrated in Fig.~\ref{figS1}(a). The asymmetric Gaussian steerability in two directions varying with $n_{A}$ is denoted in Fig.~\ref{figS1}(b), and as a consequence the induced Wigner negativity on the steering mode by applying single-photon subtraction on the steered mode is quantified in Fig.~\ref{figS1}(c). Note that the effect of thermal noise on the steered mode is more significant than that on the steering mode, which is opposite to the effect of losses on two modes in the main text. As there exists a thermal barrier in the direction $\mathcal{G}^{B\rightarrow  A}$, and correspondingly, a nonzero $\mathcal{N}_{B}(\mathcal{S}_{A|B})$ can exist only when $n_{A}<0.6816$. In the opposite direction, $\mathcal{G}^{A\rightarrow  B}>0$ and thus $\mathcal{N}_{A}(\mathcal{S}_{B|A})>0$ for arbitrarily large value of thermal noise $n_{A}$. The values of remotely created Wigner negativity are quantitatively determined by the purities of initial states as given in Eq.~\ref{WN} and plotted in Fig.~\ref{figS1}(d).

\section*{Appendix C: Necessity of the additional local Gaussian transformation $R_{A|B}$}
Here, we prove that the additional local Gaussian transformation required prior to photon subtraction in Ref.~\cite{WalschaersPRL} to optimize the induced Wigner negativity is no longer needed~\textit{if and only if} the CM of two-mode Gaussian states in the standard form $\sigma_{A B, s f}$ satisfies $c_1=-c_2=c$. This makes preparation resourceful Gaussian states for the remote generation of Wigner negativity without the need for inline squeezing.

From the standard form $\sigma_{A B, s f}$, the Schur complement of $\sigma_B$ is given by
\begin{equation}
V_{A \mid  B}=\left(\begin{array}{cc}
a-c_1^{2}/b & 0 \\
0 & a-c_2^{2}/b
\end{array}\right),
\end{equation}
whose symplectic eigenvalue is $v=\sqrt{(a-c_1^{2}/b)(a-c_2^{2}/b)}$. When there exists Gaussian steering $\mathcal{G}^{B\rightarrow  A}$, the symplectic eigenvalue $v$ must be smaller than 1 \cite{Adesso15}. Without any local Gaussian transformation $R_{A|B}$ prior to the photon subtraction on mode $A$, the condition for $W_{B}\left(\beta_{B}\right)<0$ should be $\operatorname{tr}\left[V_{A \mid  B}\right]<2$. Note that every CM $\sigma_{AB}$ that corresponds to a physical quantum state has to satisfy the \textit{bona fide} condition $a-c_1^{2}/b>0$ and $a-c_2^{2}/b>0$~\cite{Adesso15}, then we have
\begin{equation}
\operatorname{Tr}\left[V_{A \mid  B}\right]=(a-\frac{c_1^{2}}{b})+(a-\frac{c_2^{2}}{b})\geqslant 2\sqrt{(a-\frac{c_1^{2}}{b})(a-\frac{c_2^{2}}{b})}=2v.
\end{equation}
The above inequality can be saturated \textit{if and only if} $c_1^{2}=c_2^{2}$. With this condition, $\operatorname{tr}\left[V_{A \mid  B}\right]<2$ is equivalent to $v<1$, i.e., the photon subtraction on mode $A$ can always generate Wigner negativity in mode $B$ as long as $\mathcal{G}^{B\rightarrow A}>0$ without any prior local Gaussian transformation. Otherwise, if $c_1^{2}\neq c_2^{2}$, then $\operatorname{tr}\left[V_{A \mid  B}\right]>2v$, which means an additional local Gaussian transformation $R_{A|B}$ is necessary to make EPR steering sufficient for remotely generating Wigner negativity.

\section*{Appendix D: The CKW-type monogamy constraint to the distribution of Wigner negativity created in the other direction}

Steering is a directional form of nonlocality, related to the Einstein ``spooky'' paradox, which is fundamentally defined differently to entanglement. In previous work~\cite{Yumonogamy}, we have derived the CKW-type monogamy inequalities for multipartite Gaussian steering in two directions. We then wonder whether the CKW-type monogamy inequality holds for the distribution of Wigner negativity created in the opposite direction of that studied in the main text as well as in (S1), i.e., $\mathcal{N}_{A}(\mathcal{L}_{B_1B_2|A})\geq \mathcal{N}_{A}(\mathcal{L}_{B_1|A})+\mathcal{N}_{A}(\mathcal{L}_{B_2|A})$.

\begin{figure}[h]
	\includegraphics[width=78mm]{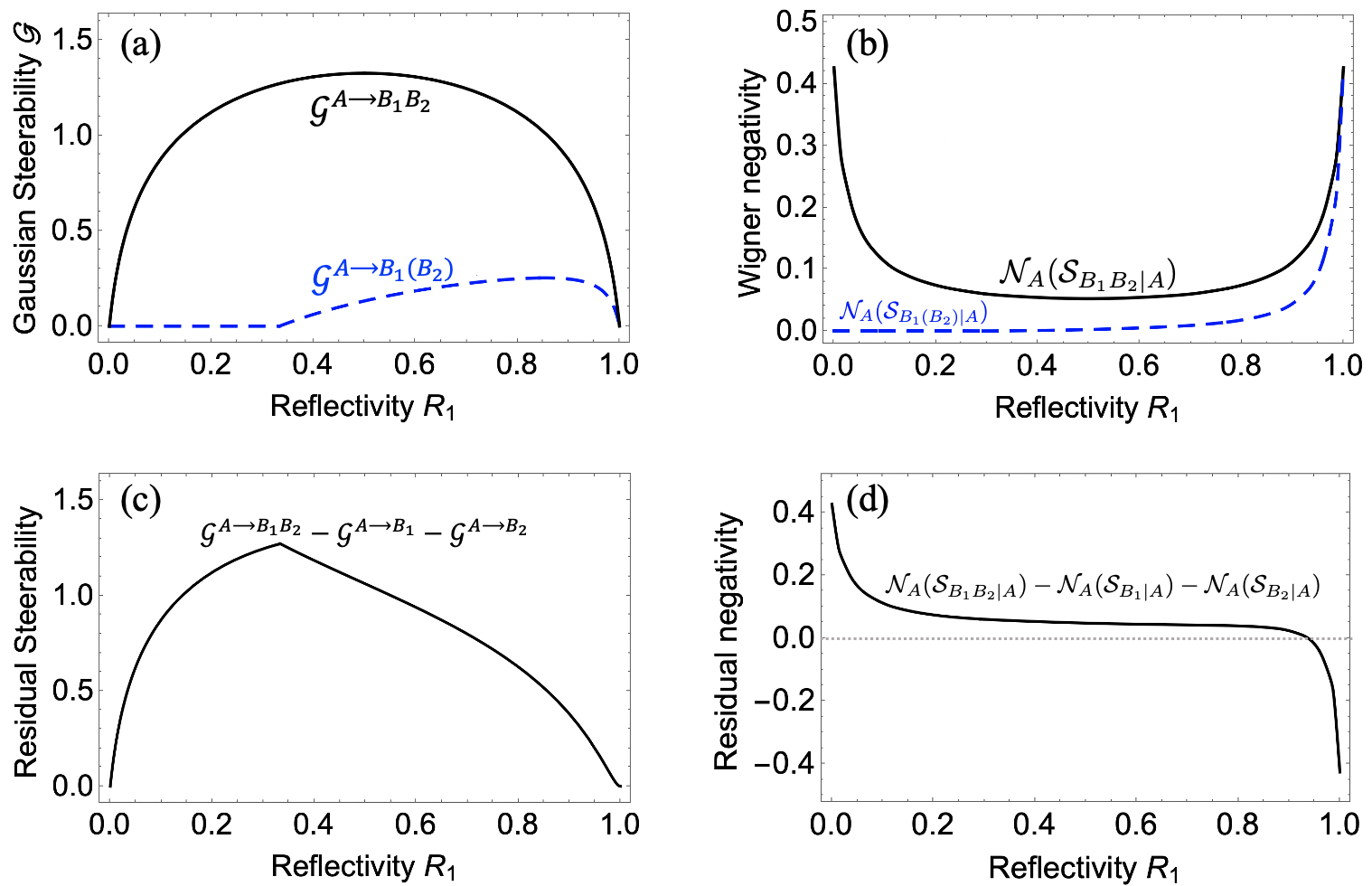}
	\caption{Scheme of a pure three-mode Gaussian state generated by the linear optical network similar to Fig.~2 in main text but with varying $R_1$ and fixed $R_2:(1-R_2)=50:50$. The input squeezing levels $r=1$ corresponding to $-8.7{\rm dB}$ quadrature noise. (a) The initial Gaussian steerabilities varying with $R_1$. (b) The corresponding transferred Wigner negativities in the steering mode $A$ by applying single-photon subtraction on the individual mode $B_1(B_2)$, or their joint $(B_1B_2)$, respectively. (c) The residual tripartite Gaussian steering $\mathcal{G}^{A\rightarrow B_1B_2}-\mathcal{G}^{A\rightarrow B_1}-\mathcal{G}^{A\rightarrow B_2}\geq 0$ which is stemming from the reverse CKW-type monogamy. (d) The residual tripartite Wigner negativity remotely created in mode $A$, $\mathcal{N}_{A}(\mathcal{S}_{B_1B_2|A})-\mathcal{N}_{A}(\mathcal{S}_{B_1|A})-\mathcal{N}_{A}(\mathcal{S}_{B_2|A})$, which is not alway larger than zero.}\label{figS2}
\end{figure}

We present a three-mode entangled Gaussian state that is similar to the case shown in Fig.~2(a) in the main text, but with the first beamsplitter being adjustable $R_1:(1-R_1)$ and the second fixed as a balanced one. In particular, when the first beamsplitter is adjusted at $R_1=1/3$, the initial Gaussian state is produced as a GHZ-like state. For the direction where mode $A$ acts as the steering party to steer the modes $B_1,~B_2$, the Gaussian steerability distributed among three modes and the corresponding Wigner negativities remotely created by a sing-photon subtraction on the individual or joint modes $B_1, B_2$ are denoted in Figs.~\ref{figS2}(a) and (b). It is clear that the two-mode and three-mode Gaussian steerability $\mathcal{G}^{A\rightarrow B_1(B_2)}>0$ and $\mathcal{G}^{A\rightarrow B_1B_2}>0$ are necessary to induce negativities $\mathcal{N}_{A}(\mathcal{S}_{B_1(B_2)|A}) $ and $\mathcal{N}_{A}(\mathcal{S}_{B_1B_2|A})$ in the Wigner functions of the steering mode $A$, respectively. Interestingly, we observe that $\mathcal{N}_{A}(\mathcal{S}_{B_1B_2|A})<\mathcal{N}_{A}(\mathcal{S}_{B_1|A})+\mathcal{N}_{A}(\mathcal{S}_{B_2|A})$ when $R_1$ approaches to $1$ (Fig.~\ref{figS2}(d)), even though the Gaussian steerability still follows the monogamy constraint, i.e., $\mathcal{G}^{A\rightarrow B_1B_2}-\mathcal{G}^{A\rightarrow B_1}-\mathcal{G}^{A\rightarrow B_2}>0$ presented in Fig.~\ref{figS2}(c). This settles an open question for the shareability of generated Wigner negativity in this direction.

\end{document}